\documentclass[conference]{IEEEtran}
\IEEEoverridecommandlockouts
\usepackage{cite}
\usepackage{amsmath,amssymb,amsfonts}
\usepackage{algorithmic}
\usepackage{graphicx}
\usepackage{textcomp}
\usepackage{xcolor}
\usepackage{caption}
\usepackage{placeins}
\usepackage{subcaption}
\usepackage{float}
\usepackage{url}
\usepackage{booktabs}
\def\BibTeX{{\rm B\kern-.05em{\sc i\kern-.025em b}\kern-.08em
    T\kern-.1667em\lower.7ex\hbox{E}\kern-.125emX}}
\begin{document}

\title{bitSMM: A bit-Serial Matrix Multiplication Accelerator
}

\author{\IEEEauthorblockN{1\textsuperscript{st} Pedro Antunes}
\IEEEauthorblockA{\textit{School of Electrical Engineering and Computer Science} \\
\textit{KTH Royal Institute of Technology}\\
Stockholm, Sweden \\
pedroa@kth.se}
\and
\IEEEauthorblockN{2\textsuperscript{nd} Artur Podobas}
\IEEEauthorblockA{\textit{School of Electrical Engineering and Computer Science} \\
\textit{KTH Royal Institute of Technology}\\
Stockholm, Sweden \\
podobas@kth.se}
}

\maketitle

\begin{abstract}
Neural-network (NN) inference is increasingly present on-board spacecraft to reduce downlink bandwidth and enable timely decision making. However, the power and reliability constraints of space missions limit the applicability of many state-of-the-art NN accelerators. This paper presents bitSMM, a bit-serial matrix multiplication accelerator built around a systolic array of bit-serial multiply--accumulate (MAC) units. The design supports runtime-configurable operand precision from 1 to 16 bits and evaluates two MAC variants: a Booth-inspired architecture and a standard binary multiplication with correction architecture. We implement bitSMM in [System]Verilog and evaluate it on an AMD ZCU104 FPGA and through ASIC physical implementation using the asap7 and nangate45 process design kits. On the FPGA, bitSMM achieves up to 19.2~GOPS and 2.973~GOPS/W, and in asap7 it achieves up to 73.22~GOPS, 552~GOPS/mm$^2$, and 40.8~GOPS/W.
\end{abstract}

\section{Introduction}
\label{sec:Introduction}
Neural networks (NNs) are expected to play an increasingly important role in space exploration. Potential applications include in-situ data analysis~\cite{nn_in_space}, on-board data compression to reduce downlink bandwidth~\cite{imp_space}, and satellite or rover navigation~\cite{navigation_dnn}. As these workloads move on-board, future spacecraft are likely to incorporate dedicated hardware to accelerate NN inference.

However, space missions impose strict constraints on computing platforms~\cite{comp_space}. In particular, radiation can induce faults, motivating radiation-tolerant designs and, in SRAM-based FPGAs, mitigation techniques such as triple modular redundancy (TMR)~\cite{space_nn,tmr_space}. In addition, limited heat dissipation and tight mass budgets require energy-efficient implementations. Conventional AI platforms, such as GPUs, are therefore often a poor fit due to their high power consumption and limited radiation tolerance.

Existing space-oriented NN accelerators primarily rely on quantized bit-parallel computations~\cite{my_survey}. While quantization can enable energy-efficient systems~\cite{quantization_survey}, current parallel architectures often lack optimizations for the extreme constraints of space environments. To date, bit-serial NN accelerators specifically for space applications remain unexplored. While binarized NNs~\cite{bnn_space} offer an alternative, their rigid quantization often results in substantial accuracy degradation that may be unacceptable for mission-critical tasks~\cite{bnn_survey}. In contrast, bit-serial architectures enable layer-specific bit-width scaling, offering a more flexible trade-off between precision and power. Furthermore, the sequential nature of bit-serial arithmetic provides a unique, yet unexamined, opportunity to integrate hardware redundancy and resiliency schemes, such as TMR, more efficiently than traditional parallel counterparts.

In this work, we present bitSMM, a bit-serial matrix multiplication accelerator designed to support space-oriented NN inference. bitSMM is built around a systolic array (SA), which is a natural fit for general matrix multiplication, the heart of most NN (e.g., CNN, MLP, and transformer)~\cite{sa_survey}. Our main contributions are:
\begin{itemize}
    \item We propose bitSMM, a bit-serial matrix multiplication accelerator.
    \item We design and evaluate two multiply--accumulate (MAC) variants---one inspired by Booth's algorithm and one based on standard binary multiplication with correction---that support runtime-configurable operand precision from 1 to 16 bits.
    \item We implement a compile-time configurable SA with an output network that reads one MAC accumulator per cycle.
    \item We implement the architecture on an AMD ZCU104 FPGA and map it to ASIC using the asap7~\cite{asap7} and nangate45~\cite{freepdk} process design kits (PDKs). On the FPGA, bitSMM achieves up to 19.2~GOPS and 2.973~GOPS/W, and in asap7 it achieves up to 73.22~GOPS, 552~GOPS/mm$^2$, and 40.8~GOPS/W.
\end{itemize}

The remainder of this paper is organized as follows. Section~\ref{sec:Section_2} provides background on binary multiplication, systolic arrays, and related work in bit-serial acceleration. Section~\ref{sec:Section_3} describes the proposed bitSMM architecture, including the bit-serial MAC variants and the SA integration and readout network. Section~\ref{sec:Section_4} presents the experimental methodology and reports FPGA and ASIC implementation results, followed by a comparison against prior work. Finally, Section~\ref{sec:Section_6} concludes the paper and outlines directions for future work.

\section{Background}
\label{sec:Section_2}
SAs are a common hardware architecture for accelerating matrix multiplication. Since matrix multiplication dominates many NN workloads in both inference and training, it is a primary target for specialized hardware. At the arithmetic level, it reduces to dot products over vectors of binary numbers. In this section, we review binary multiplication, the structure and operation of SAs, and their role in NN computation. We then summarize the state of the art (SOTA) in bit-serial and low-precision acceleration.

\subsection{Binary multiplication}
\label{sec:binary_multiplication}
In standard binary multiplication, the operands are unsigned. The product is computed by ANDing each multiplier bit with the multiplicand and shifting the partial product according to the bit position (i.e., $2^0,\,2^1,\,2^2,\,...$). Equation~\ref{eq:std_bin_mult} shows an example.
\begin{equation}
\begin{array}{r}
\phantom{+}0110_2 \\
\times\,1110_2 \\
\hline
\phantom{+}0000_2\ll 0 \\
\phantom{+}0110_2\ll 1 \\
\phantom{+}0110_2\ll 2 \\
+0110_2\ll 3 \\
\hline
01010100_2 = 84_{10}
\end{array}
\label{eq:std_bin_mult}
\end{equation}

To multiply signed numbers represented in two's complement, one can use \textbf{standard binary multiplication with correction (SBMwC)} or \textbf{Booth's algorithm}~\cite{booth_algorithm}.

In \textbf{SBMwC}, the procedure follows the unsigned case. However, at the multiplier sign bit (i.e., the most significant bit (MSb)), it applies a correction to the result. If that sign bit is 1, the multiplicand is subtracted rather than added; this subtraction is equivalent to adding the two's complement of the multiplicand. Equation~\ref{eq:signed_bin_mult} illustrates this concept with a 4-bit multiplier. Here, $0110_2=6_{10}$ and $1110_2=-2_{10}$, so the expected result is $-12_{10}$.
\begin{equation}
\begin{array}{r}
\phantom{+}00000110_2 \\
\times\,1110_2 \\
\hline
\phantom{+}00000110_2\ll 0 \\
+00000110_2\ll 1 \\
+00000110_2\ll 2 \\
((-00000110_2) \quad\text{or}\quad (+11111010_2)) \ll 3\\
\hline
11110100_2 = -12_{10}
\end{array}
\label{eq:signed_bin_mult}
\end{equation}

To explain Booth's algorithm, it is useful to recall how a binary number maps to a value. The standard representation sums powers of two at positions where the bit is one. Equation~\ref{eq:bin_to_dec} shows an example for a 4-bit number.
\begin{equation}
    0110_2 =  2^2 + 2^1 = 6_{10}
    \label{eq:bin_to_dec}
\end{equation}
An alternative (less intuitive) representation uses runs of ones: it subtracts a power of two when a 1 follows a 0 and adds a power of two when a 0 follows a 1. Equation~\ref{eq:bin_to_dec_2} shows two examples for signed numbers in two's complement.
\begin{equation}
\begin{split}
    &0110_2 =  2^3 - 2^1 = 6_{10} \\
    &1110_2 =  -2^1 = -2_{10}
    \label{eq:bin_to_dec_2}
\end{split}
\end{equation}
In \textbf{Booth's} algorithm, we scan the multiplier from the least significant bit (LSb). When we encounter the first 1, we subtract the multiplicand from the most significant bits of an accumulator. As we move to the next bit, we perform an arithmetic right shift of the combined accumulator. Next, when we encounter the first 0, we add the multiplicand. This procedure can be expressed as 2-bit recoding: at each step, we consider a pair of multiplier bits and decide whether to add, subtract, or do nothing. Table~\ref{tab:booths_control} summarizes this control logic.

\begin{table}[ht]
\centering
\caption{Booth encoding control logic. Bit pairs are taken from the multiplier starting at the LSb; $M$ denotes the multiplicand; shifts are arithmetic.}
\label{tab:booths_control}
\begin{tabular}{ll}
\toprule
\textbf{Bits} & \textbf{Action} \\
\midrule
\texttt{00} & Shift right \\
\texttt{01} & +M, shift right \\
\texttt{10} & $-M$, shift right \\
\texttt{11} & Shift right \\
\bottomrule
\end{tabular}
\end{table}

Equation~\ref{eq:booths_bin_mult} illustrates an example of Booth's algorithm.
\begin{equation}
\begin{array}{r@{\;}lr}
0110& \times\,1110  &\\
\hline
(+0000)\quad0000&\phantom{\times\,}    & \gg 1 \\
(-0110)\quad1010&\phantom{\times\,}0   & \gg 1 \\
(+0000)\quad1101&\phantom{\times\,}00  & \gg 1 \\
(+0000)\quad1110&\phantom{\times\,}100 & \gg 1 \\
\hline
\text{Result}\quad1111&\phantom{\times}0100 &
\end{array}
\label{eq:booths_bin_mult}
\end{equation}

For the first multiplier bit, we assume the previous bit is 0. Moreover, in this example, the multiplier never contains the pattern \texttt{01}; therefore, the algorithm never adds the multiplicand.

\subsection{Systolic array}
The origins of SA architectures date back to the 1970s~\cite{systolic_arrays}. Recently, SAs experienced a resurgence in interest driven by the growing demand for NN acceleration. Some examples of SOTA bit-parallel SA-based accelerators include Eyeriss, DaDianNao, Google's TPU, Samsung's TPU, and CNN-focused SA designs~\cite{eyeriss,dadiannao,google_tpu,samsung_tpu,cnn_sa}. SAs consist of multiple rows and columns of processing elements (PEs) connected in series. These PEs can be specialized to perform specific operations, such as MAC units that compute the dot product of two vectors.

\begin{figure}[ht]
    \centering
    \includegraphics[width=0.6\linewidth]{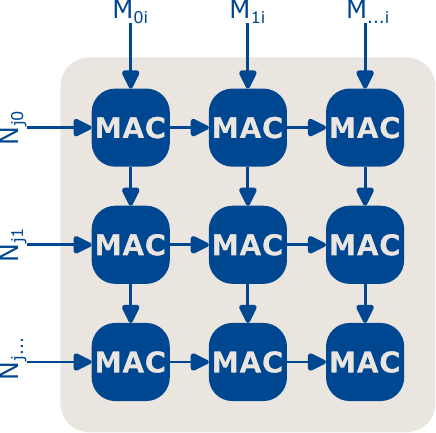}
    \caption{Generic systolic array (SA).}
    \label{fig:sa}
\end{figure}

Figure~\ref{fig:sa} depicts a generic SA architecture, where interconnected MAC units perform parallel matrix--matrix multiplication. This architecture maximizes data reuse and parallelism by streaming operands across multiple MAC units. Each MAC unit computes a specific value of the output matrix, making the SA highly efficient for the computational demands of matrix multiplication.

\subsection{Neural Networks}
The earliest implementations of NNs began with the single-layer perceptron~\cite{perceptron}. In these models, computation consists primarily of a dot product between synaptic weights and the neuron's input feature vector. Subsequent advances introduced additional layers, forming multilayer perceptrons (MLPs), whose computations can be expressed as matrix--vector multiplications. Further developments incorporated input batching and more complex layer types, enabling efficient matrix--matrix multiplication~\cite{dnn_overview}.

A typical NN comprises multiple layer types, including fully connected, convolutional, pooling, and activation layers. Among these, fully connected and convolutional layers generally dominate the computational cost. For instance, MobileNet\_v2~\cite{mobilenetv2} requires approximately $0.33\times10^9$ MAC operations, while the original Vision Transformer~\cite{vit} requires about $0.11\times10^{12}$ MAC operations.\footnote{These MAC counts were obtained using the ultralytics-thop~\cite{ultralytics_thop} Python library with models imported from torchvision~\cite{torchvision2016}.} The majority of this computation arises from matrix multiplication. Consequently, NN accelerators commonly employ SAs to efficiently accelerate these operations.

\subsection{Related Work}
\label{sec:related_Work}
Prior work exploits bit-level parallelism in several forms. Some accelerators perform bitwise operations in parallel~\cite{bitblade}, while others adopt hybrid bit-parallel/bit-serial approaches. Fully bit-serial designs stream both the multiplier and multiplicand one bit at a time into the MAC units. In many of these architectures, the MAC is integrated within a PE.

\textbf{Stripes}~\cite{stripes} investigates serial--parallel multiplication. In this design, the multiplier is streamed bit-serially, whereas the multiplicand is stored and supplied to the MAC units as 16-bit parallel values. In NN workloads, these operands correspond to activations and weights, respectively. \textbf{Dynamic Stripes}~\cite{dynamic_stripes} extends this approach by adapting activations precision at runtime.

\textbf{UNSU}~\cite{unsu} supports bit-serial weights while providing activations in parallel. It groups bits at the same position across multiple weights to index look-up tables that store partial products as functions of the activation values. These partial results are then accumulated within the PEs.

\textbf{Loom}~\cite{loom} adopts a fully bit-serial strategy. It exploits spatial parallelism by streaming one bit from 16 activations and one bit from 16 weights to each MAC concurrently. \textbf{Capra et al.}~\cite{loom_asic} later integrated a Loom-inspired design with the PULPissimo microcontroller and implemented it as an ASIC.

\textbf{BISMO}~\cite{bismo,bismo_trets} emerged around the same time as Loom. Its underlying algorithm was previously introduced by Umuroglu and Jahre~\cite{pop_bit_serial}. Although these designs are independent, their computation models are closely related. In BISMO, multiplication is decomposed into bitwise products between the multiplicand and multiplier. Each pair of bits is ANDed and then shifted according to its bit significance, which equals the sum of the bit indices. For example, for 2-bit operands the shifted bitwise products are
\begin{equation*}
\begin{array}{l}
    (mc[0]\land ml[0])\ll 0\text{, }\\
    (mc[0]\land ml[1])\ll 1\text{,}\\
    (mc[1]\land ml[0])\ll 1\text{, }\\
    (mc[1]\land ml[1])\ll 2
\end{array}
\end{equation*}
, where $mc$ denotes the multiplicand and $ml$ denotes the multiplier.
 
Without parallelism, this method requires the number of cycles given in equation~\ref{eq:bitSerial_method_1} to compute a vector dot product:
\begin{equation}
b_{mc} \times b_{ml} \times n_{values}
\label{eq:bitSerial_method_1}
\end{equation}
Here, $b_{mc}$ and $b_{ml}$ represent the bit-widths of the multiplicand and multiplier, respectively, and $n_{values}$ denotes the vector length. We use these definitions throughout the paper. Similarly to Loom, BISMO introduces intra-MAC parallelism by processing multiple operand pairs simultaneously within a MAC unit. It employs a population counter to accumulate the number of ones produced by the bitwise products of $mc[i]$ and $ml[j]$. \textbf{BISDU}~\cite{bisdu} extends this approach by integrating a BISMO-like accelerator with a RISC-V microcontroller.

These architectures~\cite{stripes,unsu,loom,bismo} primarily target convolutional and fully connected NNs. More recent work shifts toward transformer-based models. \textbf{FSSA}~\cite{fssa} introduces a SA architecture for vision transformers. In this design, each PE multiplies one activation bit with one weight bit. The weights are preloaded onto the SA, while activation bits are streamed serially. An accumulation unit then reconstructs the final outputs.

So far, these works~\cite{stripes,unsu,loom,bismo,fssa} implement conventional binary multiplication using bit-serial techniques across different architectures. More recently, \textbf{BitMoD}~\cite{bitmod} proposed an LLM accelerator that adapts Booth's algorithm by employing 3-bit encoding instead of the traditional 2-bit scheme, reducing the number of partial products.

\section{Hardware Architecture}
\label{sec:Section_3}

The proposed bit-serial matrix multiplication accelerator consists of a bit-serial MAC and a SA. The MAC computes vector dot products in a serial manner, and the SA instantiates a two-dimensional grid of these MACs interconnected through a configurable dataflow network. Its number of rows and columns is fixed at compile time. This section describes the architecture and operation of these components.

We developed the accelerator and its associated testbenches in [System]Verilog. This choice provides fine-grained control over hardware structures compared to high-level synthesis approaches while remaining technology independent.

\subsection{bitSerialMAC}

In conventional bit-parallel MAC implementations, multiplication relies on dedicated hardware multipliers, which incur significant logic overhead. In contrast, a bit-serial MAC replaces these multipliers with simple AND operations between the incoming multiplicand and multiplier bits. It accumulates the resulting partial products in a register until all vector elements have been processed and the dot product is complete. This approach substantially reduces logic complexity and area.

\begin{figure*}[ht]
\centering
\includegraphics[width=\linewidth]{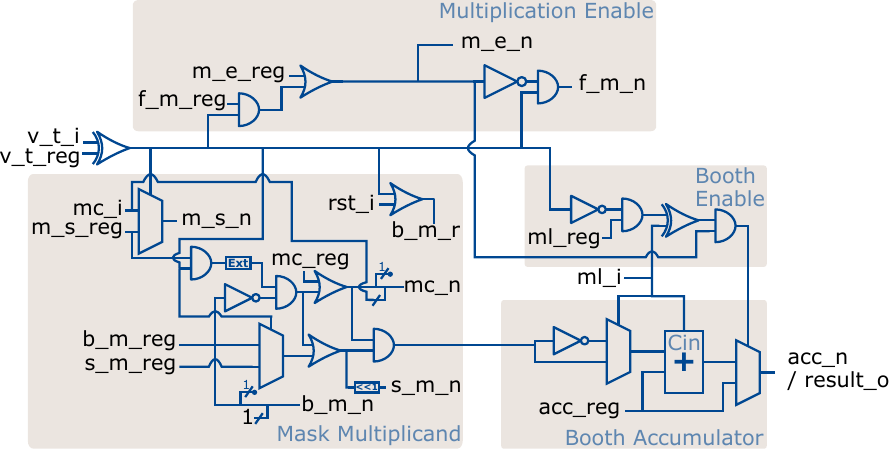}
\caption{Combinational logic of the Booth-based bit-serial MAC.}
\label{fig:bsmac_booth}
\end{figure*}

This work explores two bit-serial MAC architectures. The first architecture implements SBMwC, while the second is inspired by Booth's algorithm. Figure~\ref{fig:bsmac_booth} shows the combinational logic of the Booth-based MAC. In this diagram, signals suffixed with \texttt{\_i} denote inputs, those ending with \texttt{\_reg} originate from internal registers, \texttt{\_n} denotes register inputs, and \texttt{\_r} corresponds to reset signals. The signal \texttt{v\_t\_i} is a value toggle that changes whenever a new operand is received. This toggle replaces a cycle counter to reduce switching activity and improve power efficiency, since it changes state only when new values arrive. The inputs \texttt{mc\_i} and \texttt{ml\_i} carry the bit-serial multiplicand and multiplier, respectively.

Both MAC variants share a common multiplicand mask circuit and a multiplication-enable circuit. The maximum operand bit width is fixed at compile time; in this work, all MACs are synthesized for up to 16-bit values, consistent with common bit-parallel accelerator designs~\cite{laconic,dadiannao}. Nevertheless, their effective precision can be configured at runtime.

Within the multiplicand mask circuit, the MAC generates a bit mask by appending a leading one to a registered signal each cycle between value toggles. When it detects a new operand by XORing the toggle signal with its registered version, it copies this mask into the shift mask (\texttt{s\_m}). That shift mask isolates the active multiplicand bits that currently participate in the multiplication, allowing the next multiplicand to be loaded into the same register without corrupting the ongoing computation.

The multiplication-enable circuit detects the arrival of the first multiplicand and its corresponding multiplier bits.

In the \textbf{Booth-based MAC}, a dedicated Booth accumulator and enable circuit implement the encoding scheme described in Section~\ref{sec:binary_multiplication}. Unlike the classical formulation, the multiplicand is first sign-extended and then shifted left by one bit each cycle. The encoding itself remains unchanged: addition or subtraction is determined by examining the two most recent multiplier bits, and the Booth enable signal is asserted only when these bits differ. This design requires only a single adder.

\begin{figure}[ht]
\centering
\includegraphics[width=\linewidth]{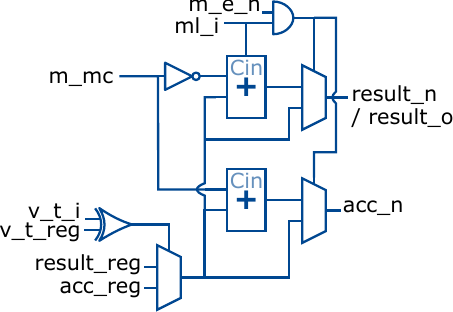}
\caption{Combinational logic of the SBMwC-based bit-serial MAC.}
\label{fig:bsmac_sbmwc}
\end{figure}

On the other hand, the \textbf{SBMwC-based MAC} requires two full adders because the MAC does not know in advance whether the current multiplier bit is the final one. If that final bit is 1, the multiplicand must be subtracted from the accumulator; otherwise, it is added. Consequently, it maintains two accumulator registers: one storing the sum and the other storing the difference with respect to the multiplicand. Figure~\ref{fig:bsmac_sbmwc} illustrates the circuit for this variant, where \texttt{m\_mc} denotes the masked multiplicand produced by the multiplicand mask unit.


In both architectures, multiplication begins by streaming the multiplicand $b_{\text{max}}$ cycles ahead of the multiplier, where
\begin{equation}
b_{\text{max}} = \max(b_{mc}, b_{ml}).
\end{equation}
During operation, each MAC receives the multiplier bits for the current multiplication concurrently with the multiplicand bits for the subsequent multiplication. Specifically, the multiplier is streamed LSb first, while the multiplicand is streamed MSb first.

Without parallelism, the number of cycles required to compute a vector dot product is
\begin{equation}
(n_{\text{values}} + 1) \times b_{\text{max}},
\label{eq:bitSerial_method_2}
\end{equation}
where $n_{\text{values}}$ denotes the number of vector elements.

Compared to prior bit-serial accelerators that support asymmetric operand bit widths, this design requires both operands to have identical widths or to be explicitly extended. Nevertheless, in terms of cycle count, it exhibits more favorable scaling behavior. Without parallelism, it achieves lower latency for all cases where $b_{mc} > 1$ and $b_{ml} > 1$ and matches prior approaches only when $b_{mc} = b_{ml} = 2$, as shown by Equations~\ref{eq:bitSerial_method_1} and~\ref{eq:bitSerial_method_2}.

Moreover, the BISMO and Loom architectures expose parallelism at the vector-value level; that is, they compute the same bit position for multiple multiplicand--multiplier pairs concurrently within a MAC. In contrast, the architecture proposed in this paper focuses on minimizing inference time while maintaining one bit stream per MAC. We anticipate that extending this design to exploit vector-level parallelism would incur greater hardware resource overhead than these prior works. Instead, it derives its parallelism advantage from executing distinct MAC operations concurrently and does not explore parallelism within a MAC.

\subsection{bitSerialSA}

The bit-serial SA (\textit{bitSerialSA}) is built from the \textit{bitSerialMAC} units described above. Its dimensions are configurable at compile time, enabling architectural exploration through script-generated Verilog integration with VeriSnip~\cite{verisnip}.

\begin{figure}[ht]
\centering
\includegraphics[width=\linewidth]{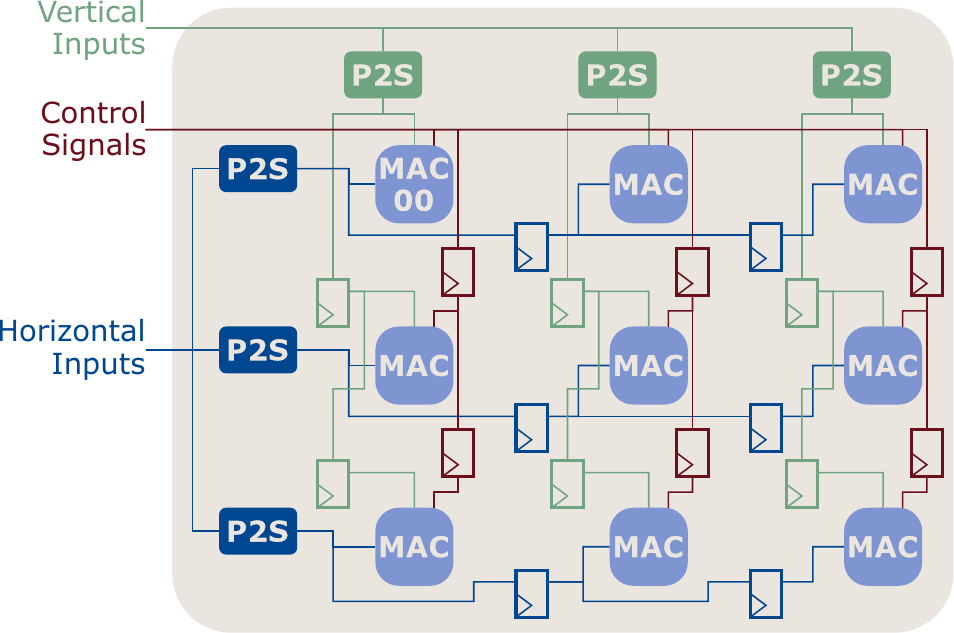}
\caption{Overall architecture of the bit-serial SA, including parallel-to-serial (P2S) converters, MAC grid, and data propagation registers.}
\label{fig:bssa_in}
\end{figure}

Figure~\ref{fig:bssa_in} depicts the overall SA structure. In addition to the MAC grid, it incorporates parallel-to-serial (P2S) units and pipeline registers that propagate data across the array. These P2S units convert parallel values fetched from memory into serial bit streams. Once the valid signal is asserted, each P2S stores a value internally and shifts it every cycle.

The P2S units connected to the vertical inputs emit the MSb first, as these inputs correspond to the multiplicands. Accordingly, their internal registers shift left each cycle. In contrast, those connected to the horizontal inputs output the LSb first and shift right, reflecting their role as multiplier inputs.


The SA accepts vertical and horizontal data streams with corresponding enable signals for each row and column, a configurable operand width, a global reset, and an output-read enable signal. Its single output exposes the MAC accumulator contents.

\begin{figure}[ht]
\centering
\includegraphics[width=0.9\linewidth]{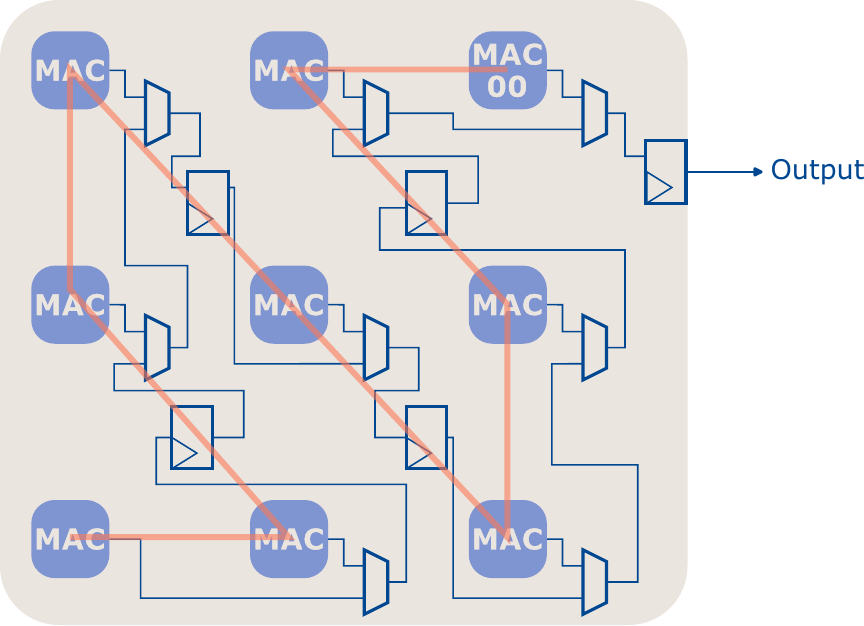}
\caption{Output readout network of the SA, showing the snake-like traversal of the read enable signal and multiplexed accumulator outputs.}
\label{fig:bssa_out}
\end{figure}

The \texttt{read\_output\_enable} signal is asserted for one cycle after a matrix multiplication completes to retrieve computation results. This signal propagates through the array in a snake-like traversal, sequentially enabling each MAC to forward its accumulator value. The traversal begins at MAC index $(0,0)$ and terminates at $(\#\text{rows}-1,\,\#\text{columns}-1)$. The output data follow the same path in reverse order, as illustrated in Figure~\ref{fig:bssa_out}.

The required number of pipeline registers is $(\#\text{rows}-1)(\#\text{columns}-1)+1$. One register resides at the final output, while the others feed two-input multiplexers. In total, $(\#\text{rows} \times \#\text{columns}) - 1$ multiplexers are used. Each multiplexer is controlled by the propagated enable signal of its corresponding MAC: when asserted, it forwards that MAC's output; otherwise, it passes the previous value along the chain.

Using this mechanism, one accumulator value is read per cycle, starting one cycle after asserting the enable signal. The total readout latency is therefore $\#\text{rows} \times \#\text{columns}$ cycles. 

The SA throughput is given in operations per second (OPS). All else equal, increasing the SA dimensions increases the maximum achievable OPS. Equation~\ref{eq:op_per_cycle} defines the number of operations per cycle for a given SA size and vector length $n$:
\begin{equation}
\mathrm{OP/cycle} = \frac{n\times Matrix\_A_{width}\times Matrix\_B_{height}}{(1+n)\times bitWidth + SA_{width}\times SA_{height}}
\label{eq:op_per_cycle}
\end{equation}

The total number of MAC operations in a matrix multiplication is $n\times Matrix\_A_{width}\times Matrix\_B_{height}$, where $n=Matrix\_A_{height}=Matrix\_B_{width}$. The total cycle count equals the computation latency in Equation~\ref{eq:bitSerial_method_2} plus the output-readout latency, which equals the number of MACs in the SA ($SA_{width}\times SA_{height}$). This design reaches peak performance as $n \rightarrow \infty$ and the input matrices match the SA dimensions. Equation~\ref{eq:op_per_cycle_peak} gives the corresponding peak throughput for any $SA_{width}$, $SA_{height}$, and bit width:
\begin{equation}
\mathrm{OP/cycle} = \frac{SA_{width}\times SA_{height}}{bitWidth}
\label{eq:op_per_cycle_peak}
\end{equation}

After implementation, multiplying operations per cycle by the system frequency yields OPS.

\section{Evaluation}
\label{sec:Section_4}

This section evaluates the \textit{bitSerialMAC} and \textit{bitSerialSA}. We first describe the experimental setup and then report results for FPGA and ASIC implementations. Finally, we compare bitSMM against prior work.

\subsection{Experimental Setup}

We developed cycle-accurate testbenches for each module to verify functional correctness. For the MAC, we exhaustively tested all multiplicand--multiplier pairs for bit widths up to 8 bits, and we tested 100 random operand pairs for bit widths between 8 and 16 bits. We also tested random vector dot products for operand widths from 1 to 16 bits and vector lengths from 1 to 1000 values. For the SA, we generated multiple \textit{bitSerialSA} topologies and evaluated matrix multiplications with varying matrix sizes (up to the SA dimensions) and varying vector lengths. For all test cases, we verified that the captured outputs matched the expected results. We executed simulations using Icarus Verilog~\cite{iverilog} version 12.0.

\begin{figure}[ht]
\centering
\includegraphics[width=0.9\linewidth]{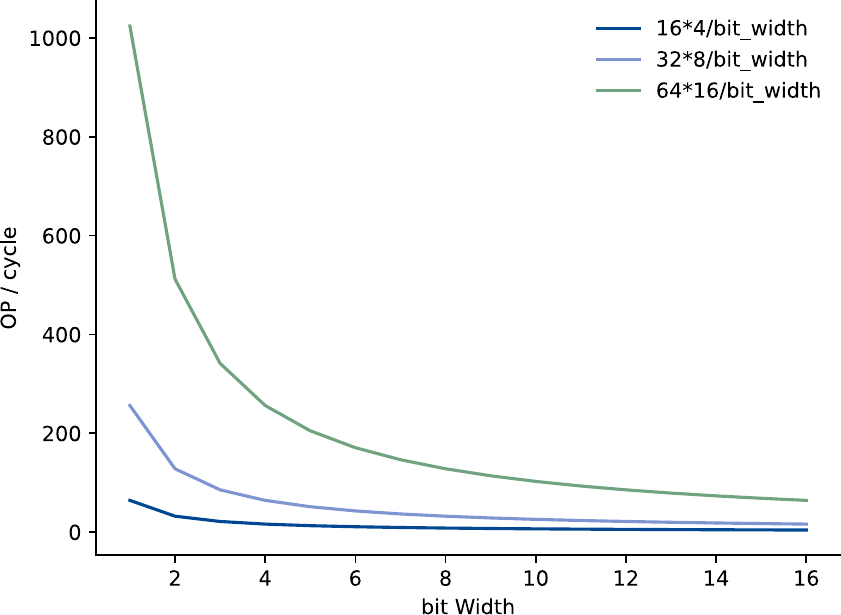}
\caption{Peak throughput (OP/cycle) for the evaluated SA topologies ($16\times 4$, $32\times 8$, and $64\times 16$) as a function of operand bit width, computed using Equation~\ref{eq:op_per_cycle_peak}.}
\label{fig:sa_peak_performance}
\end{figure}

We implemented three SA topologies on FPGA and ASIC. These topologies have rectangular dimensions of $16\times 4$, $32\times 8$, and $64\times 16$ ($\#\text{columns}$ and $\#\text{rows}$, respectively). Using Equation~\ref{eq:op_per_cycle_peak}, Figure~\ref{fig:sa_peak_performance} plots the peak throughput (OP/cycle) as a function of operand bit width for these SA sizes.

For \textbf{FPGA}, we target the AMD ZCU104 and use Vivado 2023.2 for synthesis and place-and-route. To integrate the design on the FPGA, we added a clock wizard to the top module and exposed all pins (except the clock pin) as external pins. We estimate total power consumption using Vivado and extract resource utilization from the corresponding utilization report. For \textbf{ASIC}, we use OpenROAD 2.0~\cite{openroad} to synthesize and physically implement the design using the nangate45 (45~nm) and asap7 standard-cell libraries. From the OpenROAD reports, we obtain maximum frequency, area, and estimated power.

By default, our implementations use Booth-based MAC units because we estimate that they consume fewer resources. Nevertheless, we also implement an SBMwC-based variant for a $16\times 4$ SA to compare against the Booth-based design.

\subsection{Results}

Table~\ref{tab:fpga_results} reports FPGA implementation results. bitSMM does not use Block RAMs or Digital Signal Processors (DSPs); instead, it relies only on lookup tables (LUTs) and flip-flops (FFs). As the SA dimensions increase, LUT and FF utilization increases accordingly. However, this scaling is not strictly linear: although a $64\times 16$ SA contains 4$\times$ more MACs than a $32\times 8$ SA (and $32\times 8$ contains 4$\times$ more MACs than $16\times 4$), the measured resource usage increases by more than 4$\times$ between successive configurations.

The power values in Table~\ref{tab:fpga_results} correspond to total estimated on-chip power (dynamic plus static). We compute throughput (OPS) using the equations underlying Figure~\ref{fig:sa_peak_performance} at 300~MHz, which is the target clock frequency of the FPGA implementation. We then compute throughput per watt (OPS/W) from the resulting OPS and the corresponding power.

As expected, the SBMwC-based SA consumes more resources than the Booth-based design. The Booth-based SA also achieves higher throughput per watt (GOPS/W) than the SBMwC-based SA. Although power consumption increases with larger arrays, throughput increases faster, and the $64\times 16$ SA provides the highest GOPS/W among the evaluated FPGA configurations.

\begin{table}[ht]
\centering
\caption{Implementation Results on AMD ZCU104 FPGA at 300~MHz.}
\label{tab:fpga_results}
\begin{tabular}{lccccc}
\toprule
\textbf{Design Version} & \textbf{LUTs} & \textbf{FFs} & \textbf{Power} & \textbf{GOPS} & \textbf{GOPS/W} \\
& & & (W) & & \\
\midrule
$16\times 4$       & 5630 & 8762 & 1.13 & 1.2 & 1.062 \\
$16\times 4$ SBMwC & 11418 & 10807 & 1.657 & 1.2 & 0.724 \\
$32\times 8$       & 29355 & 35490 & 2.125 & 4.8 & 2.259 \\
$64\times 16$      & 117836 & 155586 & 6.459 & 19.2 & 2.973 \\
\bottomrule
\end{tabular}
\end{table}

Table~\ref{tab:asic_results} reports results for ASIC physical implementation. We target 1~GHz for asap7 and 500~MHz for nangate45. The maximum achievable frequency is higher for smaller SAs; for those configurations, a higher target frequency would likely be feasible. Area and power scale proportionally with SA size. Notably, this results in a consistent throughput-per-watt across all implementations. We report both the peak throughput at the maximum frequency and the throughput at the target frequency. We compute GOPS/area and GOPS/W using the throughput at the target frequency.

\begin{table*}[ht]
\centering
\caption{Synthesis Results for bitSMM on ASIC Libraries.}
\label{tab:asic_results}
\begin{tabular}{lcccccccc}
\toprule
\textbf{Design Version} & \textbf{Tech Library} & \textbf{Max Freq.} & \textbf{Area} & \textbf{Power} & \textbf{Peak GOPS} & \textbf{GOPS} & \textbf{GOPS/Area} & \textbf{GOPS/W} \\
& & (MHz) & (mm$^2$) & (W) & (@ Max Freq.) & (@ Target Freq.)& (OPS/mm$^2$) & \\
\midrule
$16\times 4$         & asap7 (7nm)      & 1183 & 0.008 & 0.102 & 4.73  & 4  & 500 & 39.2 \\
$16\times 4$ (SBMwC) & asap7 (7nm)      & 1311 & 0.011 & 0.213 & 5.24  & 4  & 364 & 18.8 \\
$32\times 8$         & asap7 (7nm)      & 1124 & 0.029 & 0.403 & 17.98 & 16 & 552 & 39.7 \\
$64\times 16$        & asap7 (7nm)      & 1144 & 0.118 & 1.57  & 73.22 & 64 & 542 & 40.8 \\
$16\times 4$         & nangate45 (45nm) & 748  & 0.094 & 0.214 & 2.99  & 2  & 21.28 & 9.35 \\
$16\times 4$ (SBMwC) & nangate45 (45nm) & 730  & 0.131 & 0.305 & 2.92  & 2  & 15.27 & 6.56 \\
$32\times 8$         & nangate45 (45nm) & 685  & 0.378 & 0.809 & 10.96 & 8  & 21.16 & 9.89 \\
$64\times 16$        & nangate45 (45nm) & 643  & 1.484 & 3.28  & 41.15 & 32 & 21.56 & 9.76 \\
\bottomrule
\end{tabular}
\end{table*}

Table~\ref{tab:compare_results} compares our accelerator with other SOTA architectures. Since BISMO and FSSA operate at the bit level, these works report throughput in binary operations per second. A single 16-bit-by-16-bit multiplication requires $16\times 16=256$ binary operations in these models; therefore, we convert the reported results to 16-bit throughput to enable comparison with our 16-bit numbers.

While BISMO achieves higher performance than bitSMM on FPGAs, it necessitates data manipulation prior to execution. Additionally, although bitSMM exhibits a higher throughput than FSSA, the latter reports superior throughput per watt. In terms of area efficiency, FSSA achieves 40.86~GOPS/mm$^2$, while bitSMM achieves up to 552~GOPS/mm$^2$. Furthermore, while the FSSA design relies on a proprietary 28 nm standard-cell library, our design is implemented using entirely open-source libraries and tools.

\begin{table}[ht]
\centering
\caption{Comparison with SOTA}
\label{tab:compare_results}
\begin{tabular}{lccc}
\toprule
\textbf{Design} & \textbf{Imp. Platform} & \textbf{GOPS} & \textbf{GOPS/W} \\
& & & \\
\midrule
Opt. BISMO~\cite{bismo_trets} & ZU3EG on Ultra96 & 60 & 8.33 \\
Ours ($64\times 16$)          & ZU7EV on ZCU104  & 19.20 & 2.97 \\
FSSA~\cite{fssa}              & 28nm technology  & 25.75 & 258 \\
Ours ($64\times 16$)          & asap7 (7nm)      & 73.22 & 40.8 \\
\bottomrule
\end{tabular}
\end{table}

\section{Conclusion}
\label{sec:Section_6}
This paper presents bitSMM, a bit-serial matrix multiplication accelerator composed of a bit-serial SA and bit-serial MAC units. We evaluate multiple SA sizes and compare them against SOTA designs. Similar to other bit-serial architectures, bitSMM supports runtime-configurable precision; therefore, different layers (or groups of parameters) can use different bit-widths. To the best of our knowledge, this capability has not been studied in the context of space-oriented NN accelerators. On an AMD ZCU104 FPGA, bitSMM achieves up to 19.2~GOPS and 2.973~GOPS/W. In asap7, it achieves up to 73.22~GOPS, 552~GOPS/mm$^2$, and 40.8~GOPS/W. Overall, these results are competitive with prior work: compared to FSSA, bitSMM achieves higher GOPS/mm$^2$, whereas FSSA achieves higher GOPS/W; on FPGA, optimized BISMO still provides higher throughput than bitSMM.

In future work, bitSMM should be integrated into a complete NN accelerator to benchmark end-to-end workloads. A practical advantage of this architecture is that it does not require data manipulation in memory prior to multiplication. In particular, weights can be stored in memory in big-endian format, while activations (feature maps) can be streamed in little-endian format, which is commonly used by other system components. Finally, since some NN applications require higher precision than extreme quantization, it is important that the architecture scales linearly with operand bit width.

\section*{ACKNOWLEDGMENT}
This work is supported by the European Commission, with Automatics in Space Exploration (ASAP), project no. 101082633. Generative AI was used for drafting and language editing of selected passages. The authors reviewed and verified all technical content, results, and citations.

\bibliographystyle{ieeetr}
\bibliography{biblio}

\end{document}